\documentstyle[12pt,preprint,aps,floats,epsf]{revtex}
\tighten

\begin{document}

\preprint{
\noindent
\begin{minipage}[t]{6in}
\begin{flushright}
PM/98--22   \\
BUTP--98/24 \\
hep-ph/9809438\\
\today\\
\vspace*{2cm}
\end{flushright}
\end{minipage}
}

\title{
 Addendum to: ``Two Higgs Doublet Model     \\
 predictions for $\overline{B} \to X_s \gamma\,$
 in NLO QCD''~\footnote{Work 
supported in part by CNRS and Schweizerischer 
Nationalfonds}                             
}
\author{
Francesca M.\ Borzumati$^{a}$ and 
Christoph Greub$^{b}$
}
\address{
${}^{a}$ Laboratoire de Physique Math\'ematique et Th\'eorique, 
 Universit\'e Montpellier II, F--34095 Montpellier Cedex 5, France\\
}
\address{
${}^{b}$ Institut f\"ur Theoretische Physik,
 Universit\"at Bern, Sidlerstrasse 5, 3012 Bern, Switzerland
} 

\maketitle

\pacs{   }

\setlength{\parskip}{1.01ex}

Recently, several papers appeared that include different 
classes of  electroweak corrections~\cite{CM,KN,STRUMIA} to the 
process ${\rm BR}(\overline{B}\to X_s\gamma)$. 
In~\cite{STRUMIA}, corrections to the Wilson coefficients 
at the matching scale due to the top quark and the neutral 
higgs boson were calculated and found to be negligible. 
The analysis~\cite{CM} concluded that the most
appropriate value of $\alpha_{em}^{-1}$ to be used for this 
problem is the fine structure constant 
$\alpha^{-1} = 137.036$ instead of the value 
$\alpha_{em}^{-1} = 130.3\pm 2.3$ 
previously used. In~\cite{KN}, the leading 
logarithmic QED corrections of the form 
$\alpha \log ({\mu_W}/{\mu_b}) \left(
\alpha_s \log ({\mu_W}/{\mu_b}) \right)^n$ (with resummation 
in $n$) were given. 

We update our results of~\cite{BG} for the branching ratio
${\rm BR}(\overline{B}\to X_s\gamma)$
in the SM and for the exclusion contour plot 
$(\tan \beta,m_H)$ in a 2HDM of Type~II, by changing the 
value of $\alpha_{em}$ and by including the class of 
QED corrections presented in~\cite{KN}. They can be used 
to improve 
${\rm BR}(\overline{B}\to X_s\gamma)$ in any extension of the SM
which does not increase the set of effective operators relevant 
for the problem. 

In the SM, we obtain:  
\begin{equation}
{\rm BR}(\overline{B}\to X_s\gamma)
 = \left(
  3.32 \ \pm^{\,0.00}_{\,0.11} \ {\rm (\mu_b)}
       \ \pm^{\,0.00}_{\,0.08} \ {\rm (\mu_{\scriptscriptstyle W})}
       \ \pm^{\,0.26}_{\,0.25} \ {\rm (param)}
   \right) \times 10^{-4}  \,. 
\label{brsm}
\end{equation}
The bulk of the change with respect to the value 
presented in~\cite{BG} is due to the different 
value of $\alpha_{em}^{-1}$
used. In a 2HDM of Type~II, the new exclusion plot
in $(\tan \beta,m_H)$, obtained for different possible 
experimental upper bounds for 
${\rm BR}(\overline{B}\to X_s\gamma)$, is shown in Fig.~1.
\begin{figure}[t]
\begin{center} 
\epsfxsize=13.0 cm
\leavevmode
\epsfbox[125 435 500 700]{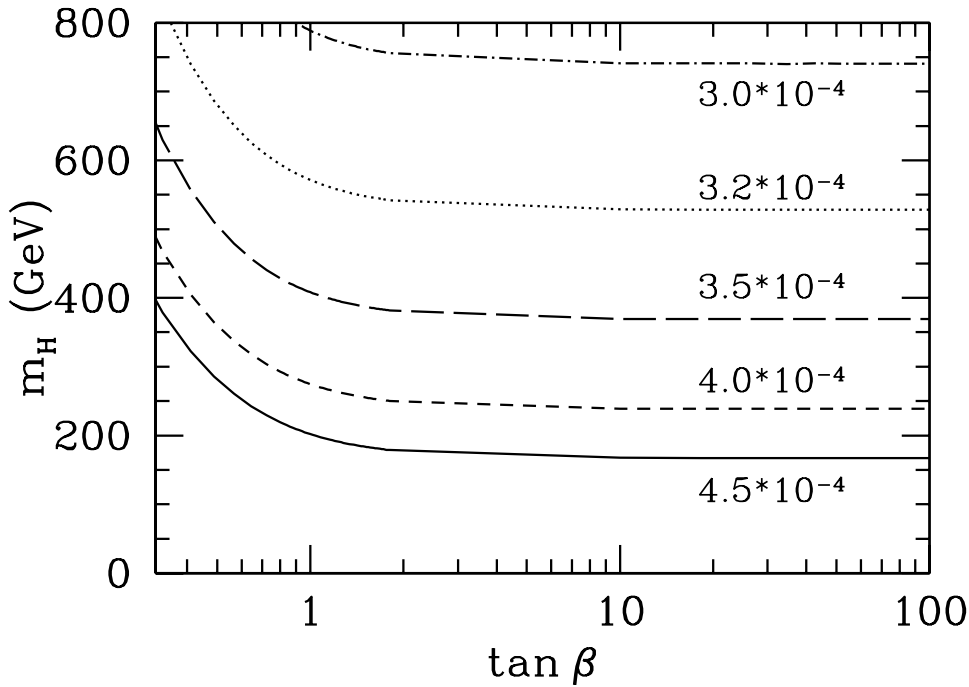}
\end{center}
\caption[f5]{\tenrm{ Contour plot in $(\tan \beta,m_H)$ obtained by 
using the NLO expression for the branching ratio 
${\rm BR}(\overline{B}\to X_s\gamma)\,$
and possible experimental upper bounds. The allowed region 
is above the corresponding curves.}}
\label{figcontour}
\end{figure}
Each curve is obtained minimizing 
${\rm BR}(\overline{B}\to X_s\gamma)/
{\rm BR}(b \to c l \nu_l)\vert_{theor}$ by varying the input 
parameters within their range of errors and the two scales 
$\mu_b$ and $\mu_{\scriptscriptstyle W}$ as described in~\cite{BG}, 
for each value of 
${\rm BR}(\overline{B}\to X_s\gamma)\vert_{exp}$ considered.

As already mentioned in~\cite{BG}, one should bear in 
mind that the error in~(\ref{brsm}) as well as that considered
to obtain the exclusion curves in Fig.~1 does not include 
all possible uncertainties in the theoretical estimate
of ${\rm BR}(\overline{B}\to X_s\gamma)$. A 
different way of handling the semileptonic 
width $\Gamma_{SL}$,
for example, retaining only the first term in the 
$\alpha_s$ expansion of $1/\Gamma_{SL}$ 
lowers the central value of 
${\rm BR}(\overline{B}\to X_s\gamma)$ from $3.32 \times 10^{-4}$
to $3.22\times 10^{-4}$ in the standard model. Similarly,
the different treatment of $1/\Gamma_{SL}$ leads to shifts
of the exclusion curves in Fig.~1 by tens of GeV for 
${\rm BR}(\overline{B}\to X_s\gamma) \sim 4 \times 10^{-4}$ 
or more for smaller 
values of ${\rm BR}(\overline{B}\to X_s\gamma)$. 

A similar effect has to be expected for additional 
electroweak corrections not included here, which 
presumably will not exceed the $2\%$ level~\cite{CM,STRUMIA}.

\end{document}